
\magnification 1200
\hsize=31pc
\vsize=55 truepc
\baselineskip=26 truept
\hfuzz=2pt
\vfuzz=4pt
\pretolerance=5000
\tolerance=5000
 \parskip=0pt plus 1pt
\parindent=16pt
\def\avd#1{\overline{#1}}

\def\P {{\cal{ P}}}

\vskip 1.truecm\noindent
\centerline{\bf CHARACTERIZATION OF CHAOS IN RANDOM MAPS}
\vskip 1.4truecm\noindent
\centerline.{V. Loreto$^{1}$,
 G. Paladin$^{2,3}$, M. Pasquini$^{1,3}$  and  A. Vulpiani$^{1}$}
\vskip .4truecm
\centerline{\it $^{1}$Dipartimento di Fisica,  Universit\`a La Sapienza}
\centerline{\it I-00185 P.le Moro 5, I-00185 Roma Italy}
\vskip .4truecm
\centerline{\it $^{2}$
NORDITA, Blegdamsvej 17, DK-2100  Copenhagen {\O} $\ $, Denmark}
\vskip .4truecm
\centerline{\it $^{3}$
 Dipartimento di Fisica,  Universit\`a dell'Aquila}
\centerline{\it I-67100 Coppito, L'Aquila, Italy}
\vskip 1.6truecm
\centerline{ABSTRACT}
\vskip .4truecm
We discuss the characterization of chaotic behaviours in random maps
both in terms of the Lyapunov exponent and of the spectral properties
of the Perron-Frobenius operator. In particular, we study a logistic map
where the control parameter is extracted at random at each time step
by considering finite dimensional approximation of the Perron-Frobenius
operator.
\hfill\break
\hfill\break
\noindent
PACS NUMBERS: 05.40.+j,05.45.+b
\vfill\eject
\vskip .4truecm
\noindent
{\bf 1. INTRODUCTION}
\medskip
The study of systems with many degrees of freedom is one of the central
problems in the field of dynamical systems [1].
In many cases, there exists a separation of the time scales,
i.e. a fast evolution superimposed to a slow one,
that allows one  to capture the dynamics in terms of simple
models given by one-dimensional random maps [2].
For instance, in the fault dynamics,
a fast evolution on small scales coexists
with a slow one on geological times [3].

In this paper we shall consider random maps where
the fast evolution is taken into account by choosing at random at
each time step a particular deterministic one dimensional map,
while the slow evolution is
given by the iteration of the map extracted time by time.
Following the ideas of Spiegel et al. [2],
we start with a system made of two variables
$x_n$ (representing the slow degrees of freedom) and
$y_n$ (the fast ones), evolving according to the deterministic
rule
$$
\left\{
\eqalign{
y_{n+1}=g_1(y_n) \cr
x_{n+1}=g_2(x_n,y_n)
}
\right.
\eqno(1)
$$
If the typical evolution time of $y_n$ is much shorter than the
corresponding time for $x_n$, such a model can be approximated in terms of
the random map
$$
x_{n+1}=f^{(i_n)}(x_n),
\eqno(2)
$$
where the integer $i_n=1,\cdots, k$ is e.g. an independent random variable,
extracted at each time step with probability $p_1,\cdots, p_n$
and $f^{(1)} ,\cdots , f^{(k)}$ are appropriate one-dimensional maps
 of the interval $[0,1]$ into itself.
This kind of random maps have been used to describe many models
such as magnetic dynamos [2], transport in fluid [4] ,
random systems [5] and on-off intermittency (abrupt switching
among regular and chaotic behaviors) [2].
Moreover, random maps exhibit many interesting
features such as scaling laws for the probability of the laminar phases [2].

The characterization of the behaviour generated by random maps
requires the extension of the tools used for the
description of deterministic chaos from a physical point of view, and
the extension of the rigorous results obtained for expanding maps
from a mathematical one. In our opinion, both these problems are
 technically non-trivial,  and worth
to be analyzed because of the  physical relevance of the phenomena
 involved.

The outline of the paper is the following.
In sect. II we introduce the basic concepts which enter in the
characterization of random maps. In particular, we discuss
the {\it average probability distribution} which is the analogue
of the invariant measure for deterministic dynamical systems,
and the so-called snapshot attractor. It is also given the definition of
the Lyapunov exponent and of complexity for random dynamical systems.
Sect. III is devoted to the Perron-Frobenius
operator for random maps. In sect. IV we study the case of a random
logistic map analyzing the three different regimes exhibited by the system for
different values of the randomness parameter $p$. We also try to get a deeper
understanding of the dynamics by using a finite dimensional
approximation of the Perron-Frobenius operator.
Finally in sect. V we draw the conclusions and indicate the possible
drawbacks of the Perron-Frobenius operator approach to the study
of random dynamical systems.

{\bf 2. BASIC TOOLS TO STUDY RANDOM MAPS}

Let us briefly summarize the basic concepts needed to
characterize random maps.

\item{1)} { \bf The average probability distribution}.

After a transient, a deterministic map evolves
on the invariant set of the dynamics (usually an attractor),
where it is possible to define an invariant probability measure.
We should now consider the probability density obtained
as a limit of the histogram of points  given by the iterations of (2),
on the `average' invariant set $I$.
It is natural to expect that such a distribution might be obtained
by means of an average over the randomness realizations.
Let us recall that for a deterministic map $f$, the invariant
 probability measure $\rho$
 is the  eigenfunction of the Perron-Frobenius operator $L$
 related to the maximum (in modulus) eigenvalue $\gamma_1=1$,
  i.e.
$$
\rho(x)=L \, \rho(x)
\eqno(3)
$$
The operator is defined as follows
$$
L \, \phi(x)=\int_I dy \, \delta (y-f(x)) \, \phi(y)=
\sum_{z=f^{-1}(x)} {\phi(z) \over |f'(z)|}
\eqno(4)
$$
A moment of reflection shows that for a random map
it is still possible to find the average probability density
by the straightforward generalization of eq.(3):
$$
\rho_{av}(x)=\avd{L} \, \rho_{av}(x)
\eqno(5)
$$
where we have introduced the (annealed) average operator
$$
\avd{L}= \sum_{j=1}^k p_j L_j
\eqno(6)
$$
and $L_j$ is the Perron-Frobenius operator associated to the
deterministic map $f^{(j)}$.

\item{2)} {\bf The snapshot attractor}.

In defining the average probability density, we have considered the
long-time evolution of a single trajectory under a `typical' realization
of the randomness.
One can also consider the probability density obtained by starting
with  a cloud of initial conditions $x_0^{(1)},\cdots,x_0^{(M)}$,
with $M$ very large, that evolve in time under {\it the same} randomness
realization of the dynamics.
There exists therefore  a instantaneous probability density $\rho_t(x)$
obtained by the histogram of the $x^{(j)}_t$ in the limit $M \to
\infty$.
Such a probability measure is usually indicated as the snapshot attractor [4].
It is generally different of the average probability density, and
asymptotically there are   the following situations:

\item{(I)} $\qquad \rho_t(x)=\delta(x-x^*)=\rho_{av}(x)$\hfill\break
where $x^*$ is a stable fixed point;

\item{(II)} $\qquad \rho_t(x)=\delta(x-x_t^*) $\hfill\break
where $x_t^*$ changes in time and $\rho_{av}(x)$ is not a delta function;

\item{(III)} $\qquad \rho_t(x) =g(x,t)$\hfill\break
where $g(x,t)$ is a non-trivial function of the space-time.

The regimes (II) and (III) can be characterized by studying other
quantities  such as Lyapunov exponents, time correlations and complexity
measures for random systems, see e.g. ref.[6].

\item{3)} {\bf The Lyapunov exponent for random maps}

In noisy systems, the Lyapunov exponent $\lambda_I$
 provides the simplest information about chaoticity and  can be computed
considering the separation of two nearby trajectories evolving in the same
realization of the random process $I(t)=i_1,i_2,...,i_t$.
It is possible to introduce the Lyapunov exponent
for random maps by considering the tangent vector evolution:
$$
z_{n+1}=\left. {df^{(i_n)} \over dx}\right|_{x_n} \, z_n
\eqno(7)
$$
so that for almost all initial conditions
$$
\lambda_I=\lim_{N \to \infty} {1\over N} \ln |z_N|.
\eqno(8)
$$
It is worth to stress that this characterization of the chaoticity of a random
dynamical
system can be misleading [6,7]. A negative value of the Lyapunov exponent
computed in such a way implies predictability ONLY IF
the realization of the randomness is known. In other more realistic
cases of trajectories initially very close and evolving under different
realization of the randomness, it can happen that after a certain
time the two trajectories will be very distant even with a negative
Lyapunov exponent $\lambda_I$.

Both in regimes $(I)$ and $(II)$, the Lyapunov exponent is negative, and in
the phase $(II)$ it also corresponds to the typical contraction rate of a
cloud  of points  toward the `jumping' snapshot attracting point
$x^*_t$, i.e. one has
$$
\sigma_t^2=\lim_{M \to \infty} {1\over M} \sum_{j=1}^M
 \left( x^{(j)}_t-{1\over M}\sum_{i=1}^Mx^{(i)}_t\right)^2
 \sim e^{-2|\lambda| t}
\eqno(9)
$$
On the other hand, we expect $\lambda >0$ in the  phase $(III)$.

\item{4)} {\bf Measures of Complexity}

We pointed out how the characterization of the chaoticity of a random
dynamical system by the Lyapunov exponent can be misleading.
On the other hand it is possible, for such a system, to introduce
a measure of complexity $K$ which better accounts for their chaotic properties
[6,7], as
$$
K \simeq h_s+\lambda_I \theta(\lambda_I),
\eqno(10)
$$
where $h_s$ is the Shannon entropy [8] of the random sequence
$I(t)$,  $\lambda_I$ is the Lyapunov exponent defined above and
$\theta$ is the Heaveside step function.
The meaning of the complexity $K$ is rather clear:
$K \log 2$ is the mean number of bits, for each iteration, necessary
to specify the sequence $x_1,...,x_t$ with a certain tolerance $\Delta$.

We stress again that a negative value of $\lambda_I$ does not implies
predictability. To illustrate this point, let us calculate
 $K$ for a system described
by a random map which exhibits the so-called {\it on-off}
intermittency [2] (see also sect.IV).
In this case one has laminar phases, i.e. $x_t \simeq 0$,  of
average length $l_L$ and
intermittent phases of average length $l_I$. It is easy to realize that for
$l_I << l_L$
$$
K \simeq  {l_I \over l_L} h_s
\eqno(11)
$$
since one has just to compute the contributions of the intermittent
bursts whose relative weight is $l_I /(l_I+l_L) \simeq l_I / l_L$.

\bigskip
{\bf 3. PERRON-FROBENIUS OPERATOR FOR RANDOM MAPS}
\medskip
A good understanding of random maps can be achieved by
studying the spectral properties of the Perron-Frobenius operator.
In practice, it is convenient to treat the problem via finite dimensional
approximations.
We have that each deterministic map $f^{(j)}(x)$ $(j=1,\cdots,k)$
defines a Perron-Frobenius operator $L_j$ via (4).
The evolution of the average density is then obtained by the
recursive relation
$$
\rho_{av}^{t}(x)= \sum_{j=1}^k p_j \,  L_j \, \rho_{av}^{t-1}(x)
\eqno(12)
$$
Requiring that $\rho_{av}$ is stationary  in time, we see that
$\rho_{av}$ is the eigenfunction of the average Perron-Frobenius operator
corresponding to the maximum eigenvalue $1$, that is
$$
\rho_{av}(x)= \sum_{j=1}^k p_j \,  L_j \, \rho_{av}(x)
\eqno(13)
$$
The Lyapunov exponents can then be obtained by a space average
over the attractor as
$$
\lambda_I=\int_I\sum_{j=1}^k p_j \ln \left| {df^{(j)}\over dx} \right|
 \, \rho_{av}(x) \, dx
\eqno(14)
$$
Moreover, the spectrum of the Perron-Frobenius operator  is
directly related to the decay rates of the time correlations also called
`resonances' of the dynamical system [9].
In particular if we consider a generic smooth  observable
$A$, the time correlation at large times decays as
$$
C_A(t)=<A(x_t) \, A(x_0)> \, - \, <A>^2 \ \sim \ e^{-\gamma t}
\eqno(15)
$$
where $\gamma >0$, $\exp(-\gamma)$ is the modulus of the
 second eigenvalue of the Perron-Frobenius operator
  and the average is a time-average,
$$
<A>=\lim_{T \to \infty} {1\over T} \sum_{i=1}^T A(x_i)
\eqno(16)
$$
For random maps $\gamma$ is given by the logarithm of the modulus
of the second eigenvalue of the average Perron-Frobenius operator
$\avd{L}$.

Beyond the average operator, it is interesting to
consider the product of random transfer operators
 operator
$$
\P_t(i_1,\cdots,i_t)=L_{i_t} \, L_{i_{t-1}} \cdots L_{i_1}
\eqno(17)
$$
that is itself a random  operator acting in an appropriate functional
space.
Roughly speaking, it can be regarded as the product of infinite
dimensional matrices whose elements are transition probabilities.
 It is therefore natural to expect that in the limit $t \to \infty$,
$(\P_t \P_t^{+})^{{1}\over {2t}}$
for almost all the randomness realizations $\{i_1,...,i_t\}$,
has a non-random spectrum, as it happens for the product
of random matrices, where the Oseledec theorem [10] holds.
This result has been proved in the mathematical literature
in the case of infinite dimensional operators acting on Banach spaces
 [11].
The operator $\P_t$ is interesting since it controls the evolution of
the instantaneous probability density $\rho_t(x)$:
$$
\rho_t(x)=\P_t \, \rho_0(x)
\eqno(18)
$$
This problem is analogous to that one of the evolution of a
tangent vector in dynamical systems, see ref.[12].
In that case the jacobian matrix plays the role of the
Perron-Frobenius operator and the tangent vector the role
of the probability density.
Note, however, that in the usual deterministic dynamical system problems
the jacobian matrix is not extracted at random, although  given
by a chaotic dynamics.

The first eigenvalue of $\P$ is equal to unity since $\rho_t(x)$
is a normalized density.  The second eigenvalue $\tilde{\gamma}_2<1$
measures the exponential rate of collapse of a generic density
$\rho_t$ toward the time-dependent eigenfunction
$\widetilde{\rho_t}$.
In other terms in an appropriate norm one has
$$
||\rho_t-\widetilde{\rho_t}|| \sim e^{t \ln|\tilde{\gamma}_2|  }
\eqno(19)
$$
It is worth stressing that in
the phase $(B)$ where the attractor is the  jumping point $x^*_t$,
$\ln|\tilde{\gamma}_2|=\lambda$ measures the contraction rate of
a cloud centered around  $x^*_t$.
On the contrary,  in the fully chaotic phase $(C)$,
$\ln|\tilde{ \lambda}_2 |$ has no particular intuitive meaning.
\bigskip
{\bf 4. TRANSITION TO CHAOS IN THE RANDOM LOGISTIC MAP}
\medskip
In this section we analyze a specific example of
random map, namely
$$
x_{t+1}=r_{i_t} \, x \, (1-x)
\eqno(20)
$$
with
$$
r_{i_t}=\left\{
\eqalign{4 \qquad {\rm with \ probability } \ p \cr
1/2 \qquad {\rm with \ probability} \ 1-p }
\right.
\eqno(21)
$$
The evolution is thus given by the random composition of a contracting
and of an expanding logistic map. For sake of simplicity the expanding
map has the control parameter set at the Ulam point ($r=4$).
Indeed, for $p=1$, it is possible to find by a well known duality argument

the invariant probability density,
$$
\rho(x)={1 \over \pi}  { 1 \over \sqrt{x(1-x)}}
\eqno(22)
$$

At varying $p$, one observes three different regimes:

\item{(1)} $p \, < {1\over 3} $.
The random map has a stable fixed point in the origin.
The average probability density $\rho_{av}(x)=\delta(x)$ and the
Lyapunov exponent is negative and it is  trivially given by
the expression
$$
\lambda^{(1)} \cdot p + \lambda^{(2)} \cdot (1-p)
\eqno(23)
$$
where $\lambda^{(1)}=4$ and $\lambda^{(2)}={1 \over 2}$
are the Lyapunov exponents of the two logistic maps (21).

\item{(2)} ${1\over 3} < \,  p \, < p_c=0.47...$.
There is a snapshot attractive fixed point $x^*_t$
and $\rho_{av}(x) \neq 0$ on all the interval $[0,1]$
while $\rho_t(x)=\delta(x-x^*_t)$.
The Lyapunov exponent is still negative and at least near the
transition the complexity is given by the expression already
cited in section 2.
$$
K \simeq {{l_{I}} \over {l_{L}}} h_s
\eqno(24)
$$
where $l_L$ and $l_I$ are respectively the average lengths of
the laminar and intermittent phases and the (24) is written in
the case $l_I << l_L$ (see fig. 1).
\item{(3)} $p > p_c$.

The random map is chaotic: the Lyapunov exponent is positive,
both $\rho_{av}(x)\neq 0$ and $\rho_t(x)\neq 0$ on all
the interval $[0,1]$.

We have tried  to get a deeper understanding on the dynamics
by considering finite dimensional approximations of the
Perron-Frobenius operator. The idea is to treat the two Perron-Frobenius
operators $L_1$ and $L_2$ related to the contracting and to the
expanding map as $N \times N$ matrices. In fact, there are two possible
strategies.

 The first one is to consider a uniform partition
of the interval $[0,1]$ made of $N$ sub-intervals. Then
we approximate $L$ by the matrix of the transition probabilities
between the partition elements. In the limit $N \to \infty$
such a matrix converges to $L$ according the Ulam conjecture [13],
 that can be proved in some cases [14-15].

The second one is to look for the matrix representation
of the operator in an appropriate  base of the
functional space of polynomials up to order $(N-1)$, i.e. the so-called
Galerkin approximation [16].

We have used both the strategies, even if it turns out that
the Ulam approximation is in practice more convenient.
 The Galerkin approximation using Legendre polynomials gives unavoidable
numerical troubles since the coefficients of the powers grow very fast
with the order $n$ of the polynomial.

We performed the numerical treatment of the Perron-Frobenius
operator as follows. For a given partition of the interval $(0,1)$
 in $N$ segments the Perron-Frobenius operator $L_{\alpha}$  $(\alpha=1,2)$
 of the map $\alpha$ is approximated by the $N \times N$ matrix
 $\hat L_{\alpha} (N)$ whose elements $[L_{\alpha} (N)]_{i,j}$
 are given by the probability to perform a transition from the interval
$j$  to the interval $i$ and $\avd{L}$ is the matrix
$$
\avd L(N) =p L_1 (N) +(1-p) L_2 (N).
\eqno(25)
$$
{}From (25) one can easily compute $\rho_{av} (x)$,
$\lambda_I$ and $\gamma$.

Fig. 2 shows $\lambda_I$ vs $p$. One can observe how the Lyapunov
exponent computed using the approximated Perron-Frobenius operator (25) is
in good agreement with the exact (numerical) results only in
the region $(III)$ and for very small values of $p$. On the contrary
the finite dimension approximation of the Perron-Frobenius operator
seems to have (if any) a very slow convergence in the region $(II)$.

Fig.3 shows $C(\tau)$ vs $\tau$ at different values of $p$. One roughly
 has for $p =p_c+ \epsilon$
$$
C(\tau) \sim e^{-\gamma(\epsilon) \tau} \tau^{-\alpha(\epsilon)}
\eqno(26)
$$
where $\gamma(\epsilon) \rightarrow 0$ for $\epsilon \rightarrow 0$.
This behaviour recalls a critical phenomenon and it is essentially
understood [2]. The approximation (25) of the Perron-Frobenius
operator gives very bad results for $\gamma$.

Fig. 4 shows $\rho_{av} (x)$ vs $x$ for two different values of $p$.
In the region (II) for
$p=p_c+\epsilon$ one has a power law behaviour
$$
\rho_{av} (x) \sim x^{-a(\epsilon)}.
\eqno(27)
$$

The approximation of the Perron-Frobenius operator is able to give  a
good approximation in the regions $(II)$ and $(III)$.

\bigskip
{\bf 5. CONCLUSIONS}
\medskip

In this paper we have discussed the dynamical characterization
of systems whose time evolution is described by random maps,
 with particular emphasis on
the notion of Lyapunov exponent and of complexity.
We have studied the application of the Perron-Frobenius operator to the
analysis of the dynamical behaviour of random dynamical systems.
The Perron-Frobenius operator can be analyzed through finite-dimensional
approximations in order to derive the invariant measure of the system,
the Lyapunov exponent and the temporal correlation functions. In
particular we have followed two ways.

The first way consists in approximating of the
Perron-Frobenius
operator as an $N \times N$ matrix whose elements are the transition
probabilities between the elements of the partition of the interval
$\left[ 0,1 \right]$ in $N$ sub-intervals.
According to the Ulam conjecture such a matrix should converge to the
Perron-Frobenius operator in the limit $ N \rightarrow \infty$.
The second approach is the so-called Galerkin approximation and
consists in the matrix representation of the operator in an appropriate
basis of the  functional space [15].
  We found out that the first strategy gives
the better results. In particular it provides a good approximation
of the average invariant measure $\rho_{av}$ (see fig.s 4a and 4b).
As for the value of the Lyapunov exponent $\lambda _I$,  the method is
suitable to describe its behaviour near the transition to chaos (from
the region $(II)$ to the region $(III)$) but the results are less
good near the first transition between the regions $(I)$ and $(II)$.
In this case, in fact, the contributions to
$$
\lambda_I= \sum p_j \int_I  \ln \vert {{df^{(j)}} \over {dx}} \vert
 \, \rho_{av}(x) \, dx
$$
could be large exactly where the Perron-Frobenius method is not
able to give an accurate approximation for $\rho_{av}$.
Fig. 2 shows that at increasing the order of approximations of the
Perron-Frobenius
operator from a $100 \times 100$ to a $200 \times 200$ matrix, there is a
tendance to obtain better and better
 approximation of the exact value of $\lambda_I$. We thus expect
that in the limit $N \rightarrow \infty$ this approach tends to
recover the real value of $\lambda_I$.
As for the temporal correlation functions the method based on the
approximation of the Perron-Frobenius operator does not work due,
probably, to a too slow convergency with $N$.

\bigskip
\noindent
{\bf Acknowledgements}
\medskip
MP acknowledges the financial support of a fellowship
 of GNSM-CNR. GP  and AV are grateful to Nordita and to the
 Niels Bohr Institute of Copenhagen  for warm hospitality.
 We thank V. Baladi for useful and interesting discussions.
\vfill\eject
\centerline {\bf Figure captions}
\bigskip
\item{Fig. 1} $x_t$ vs $t$ for the random logistic map described by eq.s
 (21) with $p=0.35$.
\medskip
\item{Fig. 2} Lyapunov exponent $\lambda_I$ vs $p$ for the logistic random
map described by eq.s(21); (a) numerical result.
Curves (b) and (c) indicate the theoretical
estimation of the Lyapunov exponent with the Ulam method with matrices
$ N \times N$ with $(b): N =200$ and $(c): N=100$.
\medskip
\item{Fig. 3} Temporal autocorrelation functions for the random logistic map
 (21) for three different values of $p$.
\medskip
\item{Fig.4} Histograms of the distribution of positions for the random
logistic map. The dot-dashed line corresponds to the theoretical estimation and
the solid line to the numerical result. (4a) $p=0.35$, (4b) $p=0.48$.
\vfill\eject
\vskip 0.4truecm
\noindent
\centerline {\bf REFERENCES}
\vskip 0.4truecm
\item{[1]} K. Kaneko, {\it Progr. Theor. Phys.}, {\bf 74} (1985), 1033.
\item{[2]} N. Platt, E.A. Spiegel and C. Tresser, {\it Phys. Rev. Lett}, {\bf
70},
(1993), 279; J.F.Heagy, N. Platt and S.M. Hammel, {\it Phys. Rev. E}, {\bf 49}
(1994),  1140.
\item{[3]} G. Lacorata and G. Paladin, {\it J. Phys. A}, {\bf 26},
(1993), 3463.
\item{[4]} L. Yu, E. Ott and Q. Chen, {\it Phys. Rev. Lett.}, {\bf 65},
(1990), 2935; {\it Physica} , {\bf 53D}, (1993), 102; S. Galluccio and
A. Vulpiani, {\it Physica A}, {\bf 212}, (1994), 75.
\item{[5]} Van Der Broeck  C. and Tel T., in {\it From Phase Transitions
to Chaos: Topics in Modern Statistical Physics}, (1992), p.227, G.Gyorgyi,
I. Kondor, L. Sasvari and T. Tel Eds (World Scientific, Singapore).
\item{[6]}  G. Paladin, M. Serva  and A. Vulpiani: {\it Phys. Rev. Lett.},
 {\bf 74}, (1995), 66.
\item{[7]} V. Loreto, G. Paladin, A. Vulpiani, {\it Phys. Rev. E}, (1995) in
press
\item{[8]} C.E. Shannon and W. Weaver,  {\it The mathematical theory
of communications}, Univ. of Illinois press, Urbana (1949).
\item{[9]}  J.P. Eckmann and D. Ruelle, {\it Rev. Mod. Phys}, {\bf 57}, (1985),
617.
\item{[10]} V.I. Oseledec, {\it Trans. Moscow. Math. Soc.}, {\bf 19},
(1968), 197.
\item{[11]} P. Thieullen,  {\it A. Inst. H. Poincar\'e} - Analyse Nonlineaire
   {\bf 4}, (1987), 49.
\item{[12]} S.A. Orszag, P.L. Sulem and I. Goldhirsch, {\it Physica D}, {\bf
 27}, (1987), 311.
\item{[13]} S.M. Ulam,       {\it A collection of mathematical problems}
        New York, London Interscience 1960
        Interscience Tracts in Pure and Applied Mathematics n.8
\item{[14]} T. Y. Li,  {\it J. of Approx. Theory}
    {\bf 17} (1976), 177.
\item{[15]} V. Baladi, S. Isola and B. Schmitt, {\it A. Inst. H.
Poincar\'e}     {\bf 62} (1995), 251.
\item{[16]}  F. Christiansen, G. Paladin and H.H. Rugh,
{\it Phys. Rev. Lett. } {\bf 65},  (1990), 2087.

\bye